\documentclass[preprint]{aastex631}

\usepackage{comment}
\usepackage{enumitem}
\usepackage{amsfonts,amsmath,amssymb}

\usepackage{xspace}

\newcommand{\thisgrb}{GRB~190530A\xspace}
\newcommand{\sw}[1]{\texttt{#1}}

\newcommand{\fermi}{{\em Fermi}\xspace}

\newcommand{\fermiT}{{T$_{\rm 0}$}\xspace}
\newcommand{\keV}{{\rm keV}\xspace}
\newcommand{\meV}{{\rm MeV}\xspace}

\newcommand{\s}{{\rm ~s}\xspace}

\hypersetup{linkcolor=red,citecolor=cyan,filecolor=blue,urlcolor=magenta}
\usepackage{subfigure}
\shorttitle{GRB 109530A: test to LIV}
\shortauthors{Liang et al.}

\def\P{{\rm P}}
\def\lappeq{\mathrel{\rlap{\raise.5ex\hbox{$<$}}{\lower.5ex\hbox{$\sim$}}}}

\begin{document}

\title{Spectral evolution responsible for the transition from positive lags to negative lags in Gamma-ray Bursts}

\correspondingauthor{Rui-Jing Lu}
\email{luruijing@gxu.edu.cn}

\author{Wen-Qiang Liang}
\author{Rui-Jing Lu}
\author{Cheng-Feng Peng}
\author{Wen-Hao Chen}

\affiliation{Laboratory for Relativistic Astrophysics, Department of Physics, Guangxi University, Nanning 530004, People’s Republic of China}

\begin{abstract}
It was well known that most of gamma-ray bursts (GRBs) are dominated by positive spectral lags, while a small fraction of GRBs show negative lags. However,  Wei et al. firstly identified a well-defined transition from positive lags to negative lags in GRB 160625B, and then got robust limits on possible violation of Lorentz Invariance (LIV) based on the observation. Recently, such a transition has been found in three different emission episodes in \thisgrb by Gunapati et al., which provides us a great opportunity to investigate whether the transition results from LIV-induced observed spectral lags. Our analysis shows that the LIV model can not be compatible with  the current observations, whereas, only the spectral evolution induced spectral lags could responsible for the transition. So, spectral evolution can also explain the positive to negative lag in GRB 190530A. 
\end{abstract}


\section{INTRODUCTION} \label{sec:intro}
GRBs are the most energetic electromagnetic explosions in the Universe. The temporal structure of prompt gamma-ray emission exhibits diverse morphologies (\citealp{Fishman_GJ-1995-Meegan_CA-ARA&A.33.415F}), which can vary from a single smooth large pulse to an extremely complex light curve
with many erratic overlapping pulses (\citealp{Peer_A-2015-AdAst2015E.22P}). It is also generally believed that an individual shock episode gives rise to a pulse, in which the radiation spectrum evolves uniformly \citep{Lu2018}, which suggests that pulses are fundamental units of GRB radiation.  

The spectral lag, referring to the difference of the arrival time for different energy photons, is a common feature of pulses in GRBs \citep{Norri86,Cheng95,Band97,Norris00}. Early in the BATSE era, it was found that most of the GRBs pulses are dominated by the positive lags (conventionally defined as positive lags when the low energy photons lag behind the high energy photons), and a small fraction of pulses show negative lags, or negligible spectral lags (e.g., \citealt{Norris00,Yi06,Lu2018}). Generally, the bursts with long-wide pulses tend to have long lags \citep{Norris05}. 

Despite of decades of research, the true physical mechanism responsible for the spectral lag remains incompletely resolved to date. It was suggested that the curvature effect of the spherical fireball is a plausible explanation for the spectral lag (e.g., \citealt{Shen05,Lu06,Shenoy13}). In this scenario, the emission from the spherical shells at progressively higher latitudes with respect to the observer's line of sight is progressively delayed due to the weaker Doppler-boosting effect, such that the light curves of low energy radiation would be delayed and peak at later times. But the main difficulty of the curvature effect model is that the flux levels at different energy bands are particularly lower than those observed \citep{Zhang09}. Then \cite{Uhm16} proposed a more reasonable physical model to interpret the observed positive lags, in which, apart from the curvature effect, more physical mechanisms, such as the intrinsic curved spectral shape, the evolution of the magnetic field strength, and the rapid bulk acceleration of the emission zone, have been considered. Even so, the aforementioned theories can not explain the rarely observed negative lags. 

There are two different spectral evolutionary patterns within individual pulses: the peak energy ($E_{\rm p}$) of a $\nu f_{\rm \nu}$ spectrum tracking the flux (``track") and the $E_{\rm p}$ decreasing monotonically while the flux rises and falls (``hard-to-soft"; \cite{Wheaton73,Golenetskii83,Norri86,Lu10}). It is also discussed that the time evolution of the $E_{\rm p}$  may cause observed spectral lag \citep{Ukwatta12}. Furthermore, systematical investigations show that the spectral lags are correlated to the spectral evolutions as well as their corresponding spectral shapes \citep{Lu2018, Du2019}: The positive type of spectral lags can occur in either a ``hard-to-soft" pulse or a tracking pulse with $k_{\rm E}<-0.2$\footnote{In the paper of \cite{Lu2018}, to identity the main trend of spectral evolution, they had performed a linear fit to the data sets ($\log(t),\log E_{\rm p}$ ) measured in a pulse, i.e., $\log E_{\rm p} = k_E\log(t)+b$, where $t$ is the observer time relative to the beginning of the pulse, thus a steeper slope (i.e., larger $|k_E|$) indicates a faster softening, and a value of $k_{\rm E}\simeq0$ means no clear spectral evolution.}(``hard-to-soft"-dominated tracking pulse), whereas a tracking pulse with $k_{\rm E}>0.1$ (``soft-to-hard"-dominated tracking pulse) takes on the negative type of spectral lags, and no spectral lag was identified in a tracking pulse with $k_{\rm E}\simeq0$. And then the theoretical studies confirmed the observational features based on the simple physical model invoking synchrotron radiation emitted from a bulk-accelerating outflow at a large distance from the central engine \citep{Uhm2018}.

Another possible explanation for such negative lags is possibly provided by cosmological LIV, which rises in various frameworks and theories of quantum gravity (QG). In the context of QG theories, LIV occurs around the Planck energy scale ($E_{\mathrm{P} 1}=\sqrt{\hbar c^{5} / G} \simeq 1.22 \times 10^{19} \mathrm{GeV}$) \citep{Amelino1998}. IF the  LIV does take place, then among photons emitted simultaneously from the same source, the photons with high-energy will arrive at the Earth slowly, i.e., the arrival time delay of high-energy photons. Therefore, GRBs are well-suited for constraining LIV due to their large cosmological distance, small variability time scale, and very high energy photons. Using the delay time of photons in an individual GRBs, several authors (e.g., \cite{Amelino1998,Rodr06,Abdo09a, Abdo09b}) have gotten various limits on LIV.

However, these limits face an important challenge that comes from the fact that the observed time delay is a combination between the intrinsic time delay due to the emission mechanism of the source and the possible LIV-induced time delay. The problem of the intrinsic time delay is more complicated, since it requires precise knowledge of the physics of the source. To alleviate this problem, several authors (e.g., \citealt{Ellis06,Bolmont08,Biesiada09,Pan15}) have proposed some methods, which are based on the idea that the possible LIV-induced time delay will only depend on the redshift, while the intrinsic time delay will not. Thus GRBs with different redshifts will have the same intrinsic time delay (for more detailed imformation please referered to \cite{Addazi22} for a review). However whether there are the same intrinsic time delay in different emission episode in a GRB or not is yet known. 

Interestingly, when investigated the evolution features in the spectral lags in an extremely bright of GRB 160625B based on the cross-correlation function (CCF) method, \cite{Wei2017} had firstly identified a well-defined transition from positive lags to negative lags. Thus by employing a power-law function to describe the intrinsic positive time delay quantitatively, and the negative time delay in the high energy band being induced by the possible LIV effect, they have obtained a 1$\sigma$ lower limit on LIV: $E_{\rm {QG,1}} \geq 0.5\times10^{16}$ GeV for linear scale, and $E_{\rm {QG,2}} \geq 1.4\times10^{7}$ GeV for the quadratic LIV energy scale. But one should keep this in mind that, there is not any physics behind this phenomenological functional form. Moreover, Whether there are the same parameters of this function in different emission episode in a GRB or not needs to be furtherly comfirmed with observations.

\thisgrb simultaneously triggered GBM \citep{2009ApJ...702..791M} and LAT \citep{2009ApJ...697.1071A} onboard \fermi at 10:19:08 UT on May 30, 2019 (\fermiT). The best on-ground \fermi GBM position is RA, DEC = 116.9, 34.0 degrees (J2000) with an uncertainty radius of 1$^{\circ}$ \citep{Fermi2019,Longo2019}. The GBM light curve comprises of three bright emission peaks with a duration of 18.4 s (in 50 - 300 \keV energy channel).  

Recently, \cite{Gupta2022} have performed in detailed the observed spectral lag analysis for \thisgrb by the means of the cross-correlation analysis, and found that there exists a turn-over in the spectral lag (see Fig.(A2) in their paper) in the second, third episodes, and the total emission interval, respectively, the same feature seen in GRB 160625A \citep{Wei2017}. The observation provides us a great opportunity to investigate whether the power-law function mentioned above could share the same parameters (both intercept and slope) in different individual pulses in a GRB, and whether the turn-over is the signature of the violations of Lorentz invariance, and If not, there are another possible explanations for this turn-over feature.  

The outline of this paper is as follows. We tentatively constraint on the violation of Lorentz invariance based on the same method as \citep{Wei2017} in Section \ref{sec:LIV}. And then, for a comparison, spectral evolution induced observed spectral lags is introduced for another possible explanation in Section \ref{sec:spectral evolution}. Finally, our discussion and summary are given in the last section. 

\section{Case of LIV-induced observed spectral lags} \label{sec:LIV}

If LIV-induced time-lags are considered, the observed spectral lags between high- and low-energy photons emitted from GRBs can be expressed as:
\begin{equation}
\Delta t_{\rm {obs}} = \Delta t_{\rm {int}} + \Delta t_{\rm {LIV}},
\label{eq:sum}
\end{equation}
where \(\Delta t_{\rm {int}}\) is the intrinsic time-lag between the emission of photon of a particular energy and the lowest energy photon from the GRB, and $\Delta t_{\rm {LIV}}$ is the LIV-induced time lags. 

Based on the observations that the observed spectral lags of most GRBs have a positive energy dependence (see \cite{Shao2017} ; \cite{Lu2018}), \cite{Wei2017} firstly invoked a power-law functon to describe the feature of this intrinsic time-lag quantitatively. In this paper, we follow the model, thus the intrinsic time-lag can be written as following: 
\begin{equation}
\Delta t_{\rm {int}}  = \tau \left[\left(\frac{E_{h}}{keV}\right)^\alpha -  \left(\frac{E_{l}}{keV}\right)^\alpha \right] , 
\label{eq:int}
\end{equation}
\noindent where $E_h$ and $E_l$ are the energies of high- and low-energy photons, whereas  $\tau$ and $\alpha$ are free parameters in the following analysis. 

The LIV-induced time lags, occurring at a considerably higher energy (closed to Planck energy scale) and can be written as (detailed information please referred to \cite{Amelino1998,Jacob2008}):
\begin{equation}
\label{eq:liv}
\Delta t_{\rm {LIV}}=t_\mathrm{h}-t_\mathrm{l} \simeq s_\mathrm{\pm}\frac{n+1}{2}\,\frac{E_h^n - E_l^n}{\mathrm{H}_\mathrm{0} E_{\rm {QG,n}}^n}\ \kappa_n(z),
\end{equation}
where $E_{\rm {QG,n}}$ is the $n$th-order quantum gravity energy, above which Lorentz violation kicks in; $s_\mathrm{\pm}$ is the theoretically relevant factor, $s$ =+1 ( -1 ) represents the high-energy photons are slower (faster) than the low-energy ones. $\kappa_n(z)$ is a parameter depending on the distance of the source \citep{Jacob2008}, 
\begin{equation}
\kappa_n(z) =\int_{0}^{z} \frac{(1+z^\prime)^n dz^\prime}{\sqrt{\Omega_M (1+z^\prime)^3 + \Omega_\Lambda}}.
\label{eq:kn}
\end{equation}
In this paper, we take the cosmological parameters values from Planck results (\cite{Planck:2018vyg};\cite{Zyla:2020zbs}): $\mathrm{H}_\mathrm{0} = 67.4\mathrm{km\,s}^{-1}\,\mathrm{Mpc}^{-1}$, $\Omega_m = 0.315$ and $\Omega_\Lambda = 0.685$.

There are four-time intervals in \thisgrb, covering the total emission interval (time interval: (-1, 20) s relative to GBM trigger), the first (time interval (-1, 5) s), second (time interval (7, 12) s) and third (time interval (12, 20) s) episodes. With the CCF analysis method, \cite{Gupta2022} have identified that there exists the turn-over feature, as seen in GRB 160625A \citep{Wei2017}, in three different emission episodes (the second, third episodes, and the total emission interval, please referred to Fig.(A2) of their paper) of \thisgrb.

Therefore, similar to the analysis in \cite{Wei2017}, when performing the constraint on LIV model parameters from observation of energy-dependent time delay measured in the three episodes with the turn-over features, we only consider the leading dominant term in Eq. (\ref{eq:liv}), either linear ($n=1$) or quadratic ($n=2$). As only linear $n=1$ or quadratic $n=2$ modifications are of interest for experimental searches taking into account the sensitivity of current detectors \citep{Bolmont08}. 

\begin{figure}
\centering
\includegraphics[scale=0.33]{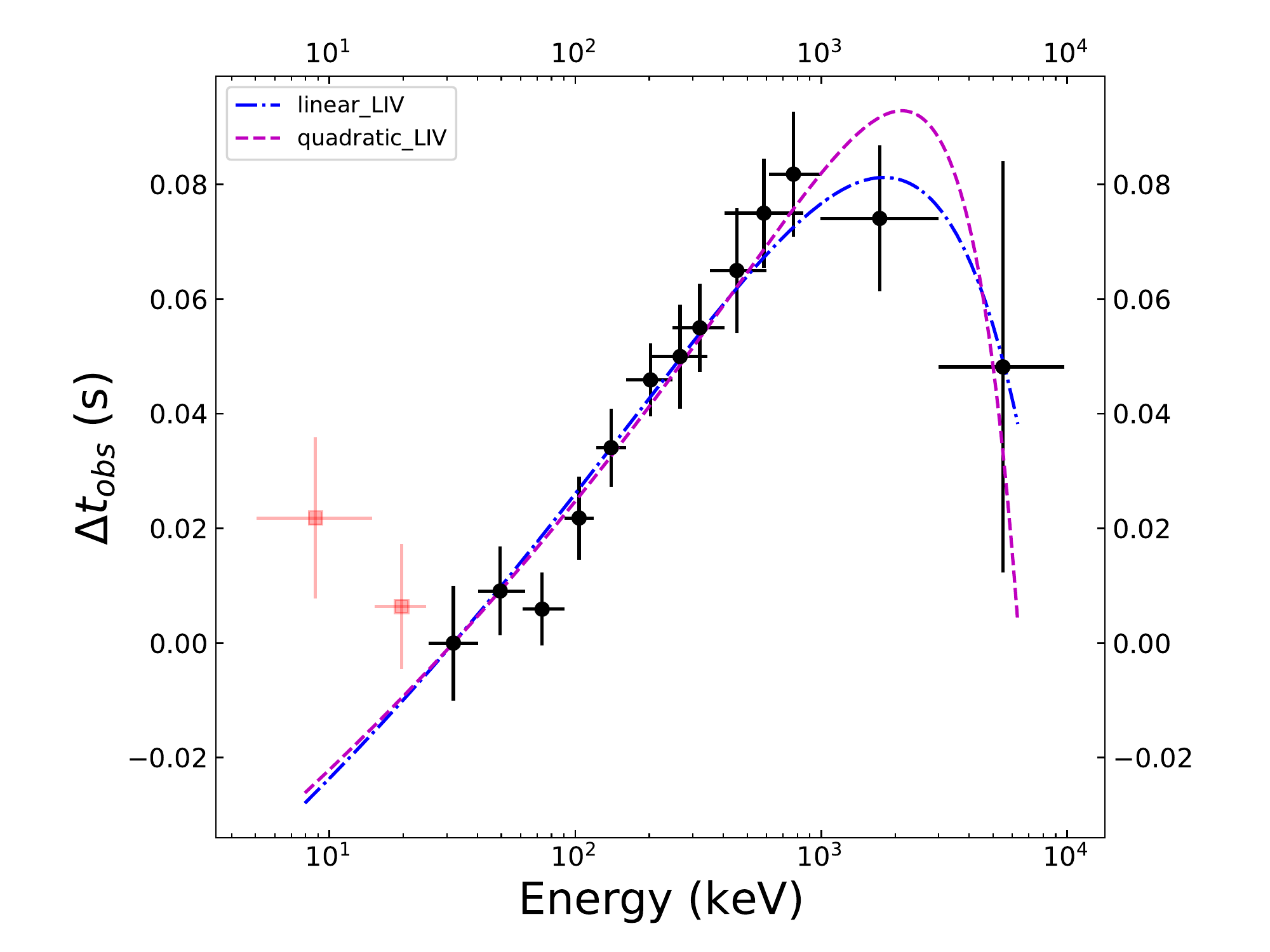}
\includegraphics[scale=0.28]{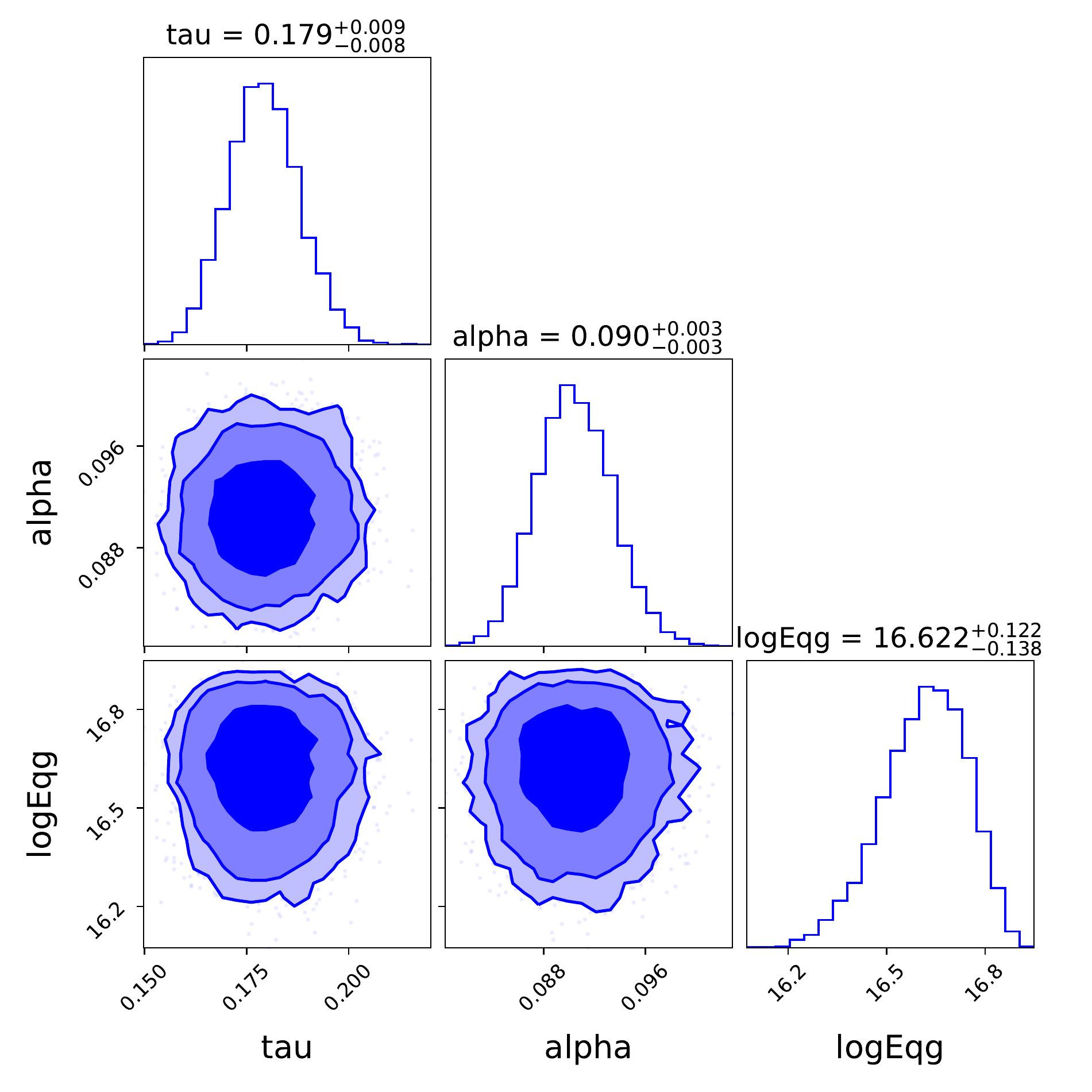}
\includegraphics[scale=0.28]{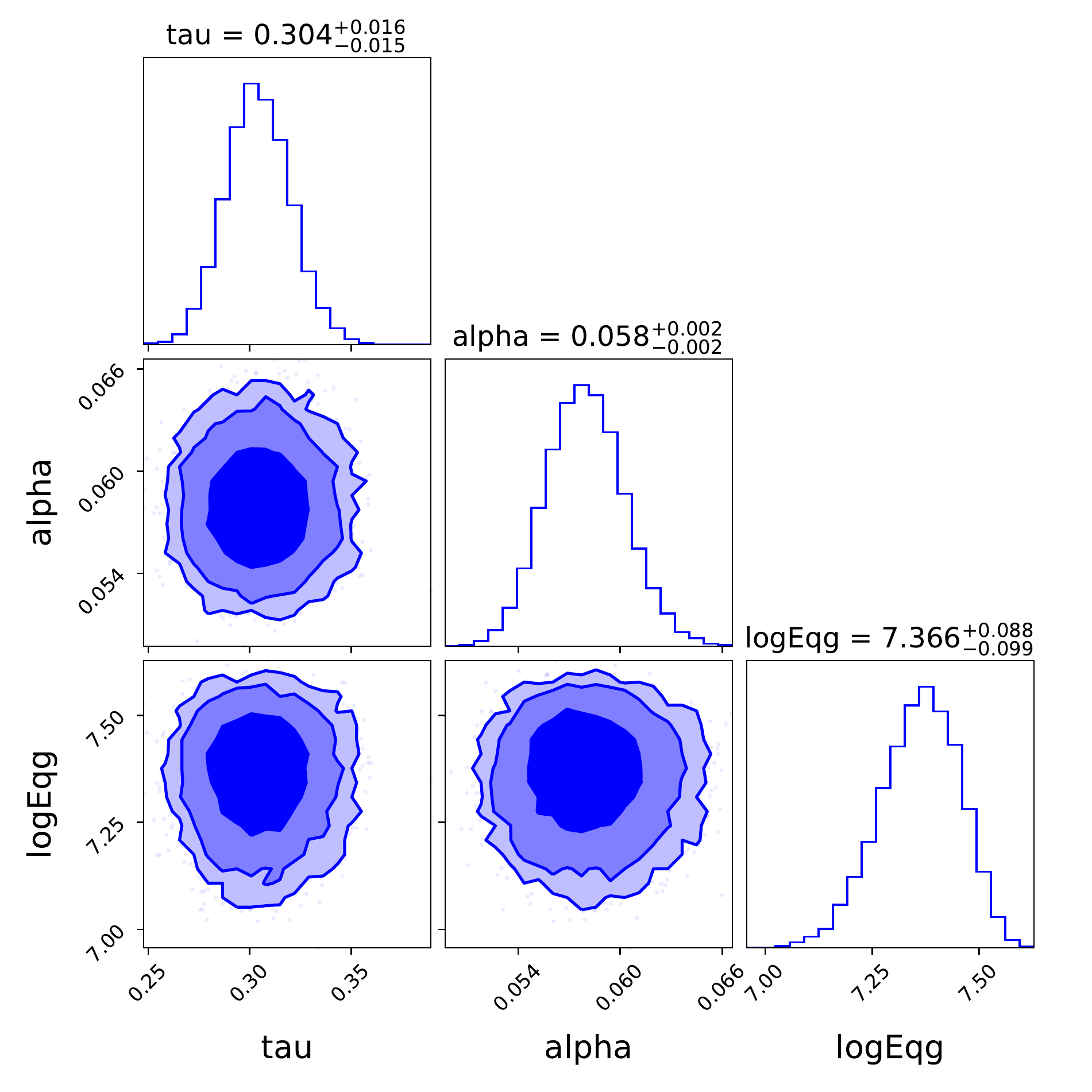}
\caption{Demonstrations (left) of the compatibility of the best fit LIV models for $n=1$ and $n=2$ to the observed spectral lags in the second emission episode of GRB 190530A (data from Gupta et al.(2022)). Noted that the spectral lag relative to the lowest channel, and the red square points are excluded from the fitting. Variational inference is adoped for the LIV models parameter estimation based on pymc3 sampler, and the corresponding credible intervals of the LIV models parameters for $n=1$ (middle panel) and $n=2$ (right panel) are plotted using the {\tt corner.py} module, in which 68\%, 90\% and 95\% credible intervals of parameters are also plotted in different colors. }
\label{pusle2}
\end{figure}

\begin{figure}
\centering
\includegraphics[scale=0.33]{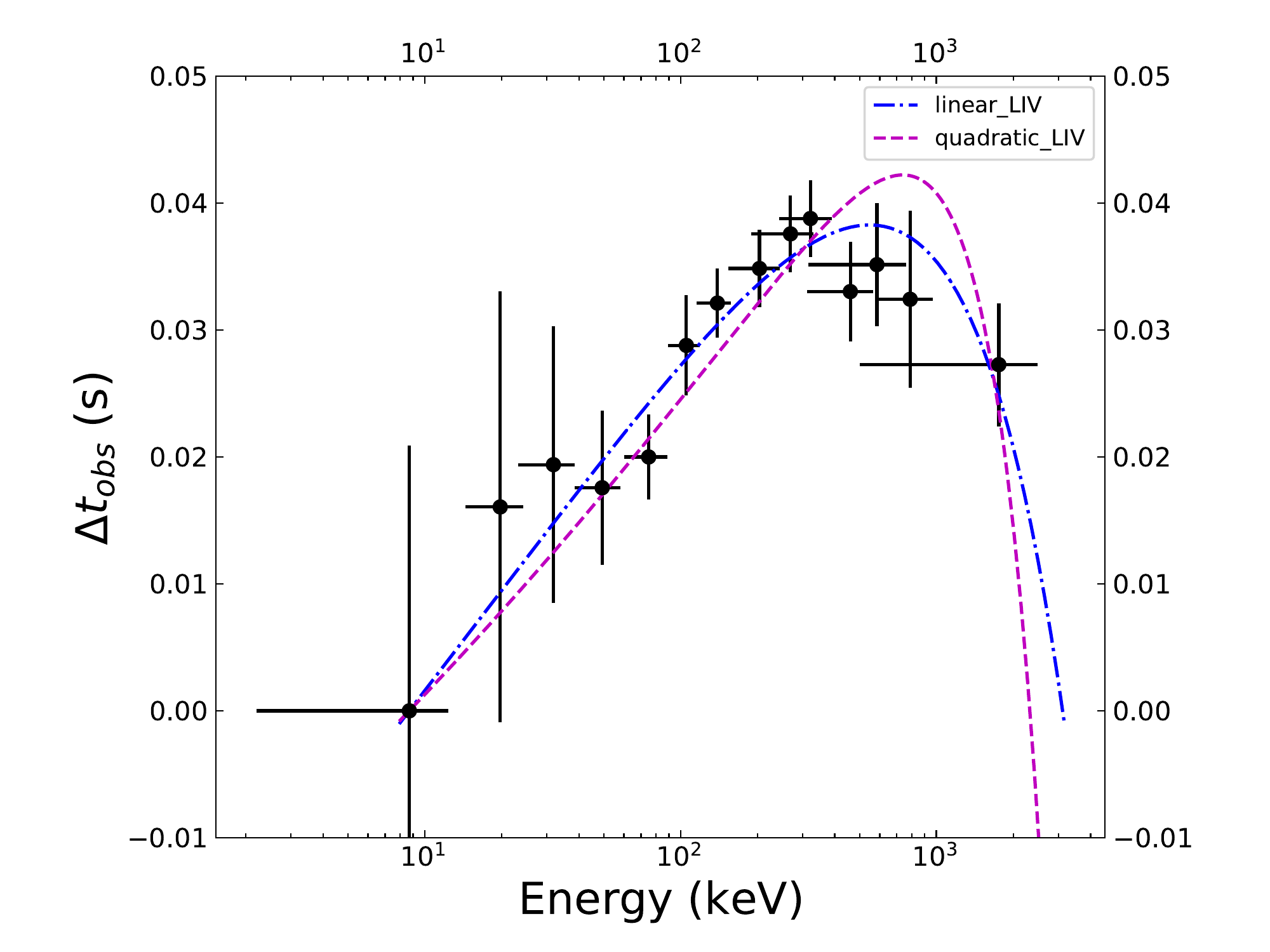}
\includegraphics[scale=0.28]{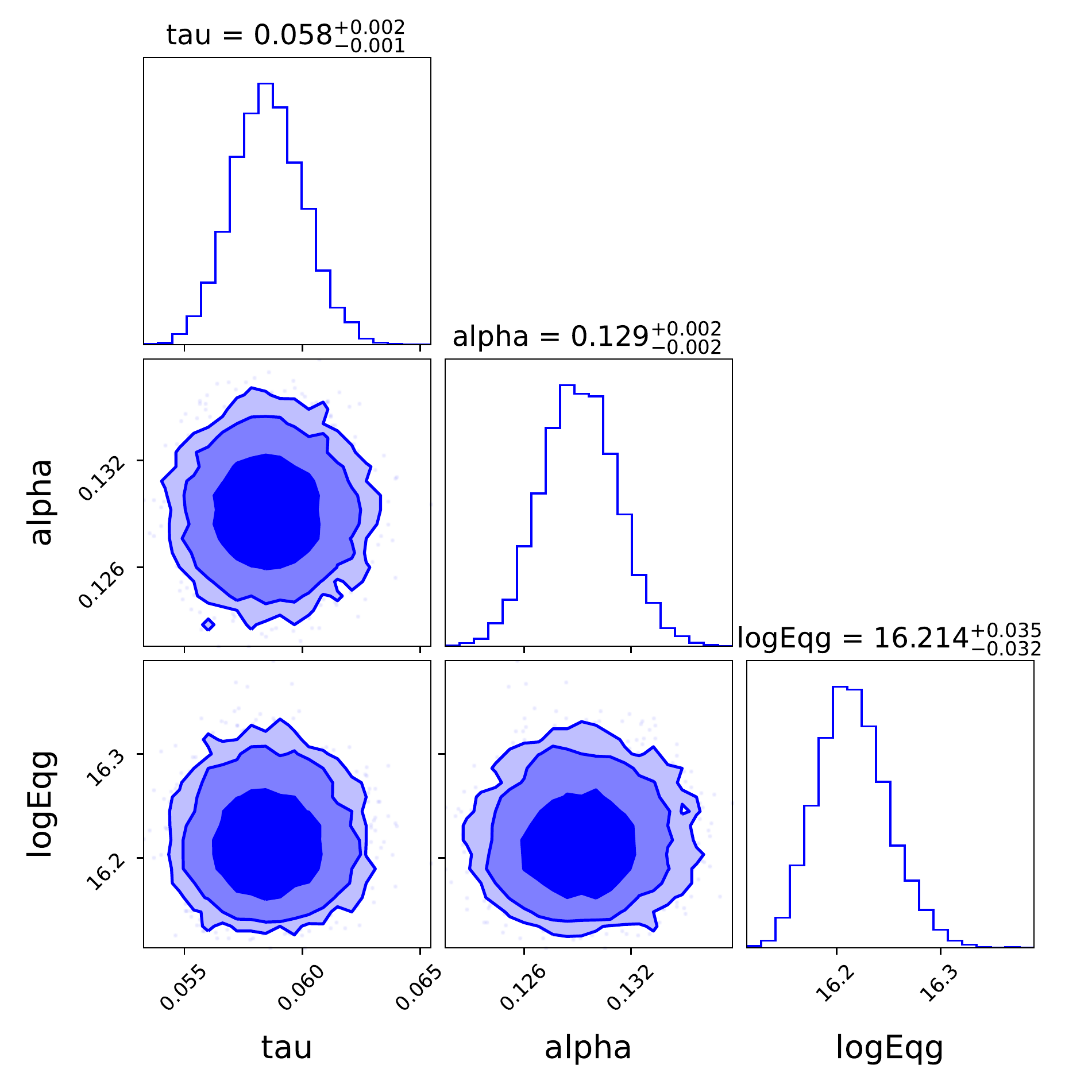}
\includegraphics[scale=0.28]{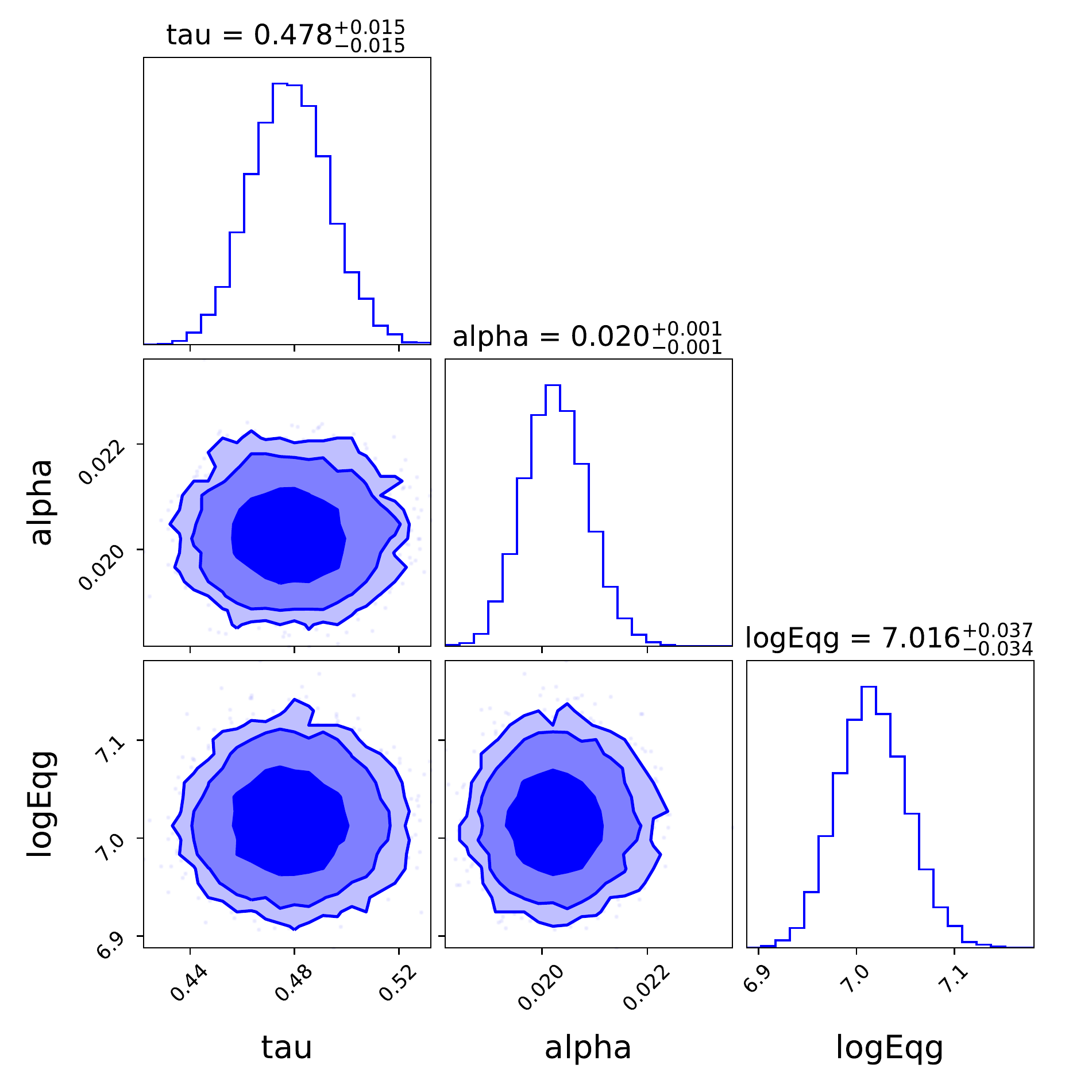}
\caption{The same as Fig.\ref{pusle2} but for the observed spectral lags in the third emission episode of GRB 190530A. }
\label{pusle3}
\end{figure}

\begin{figure}
\centering
\includegraphics[scale=0.33]{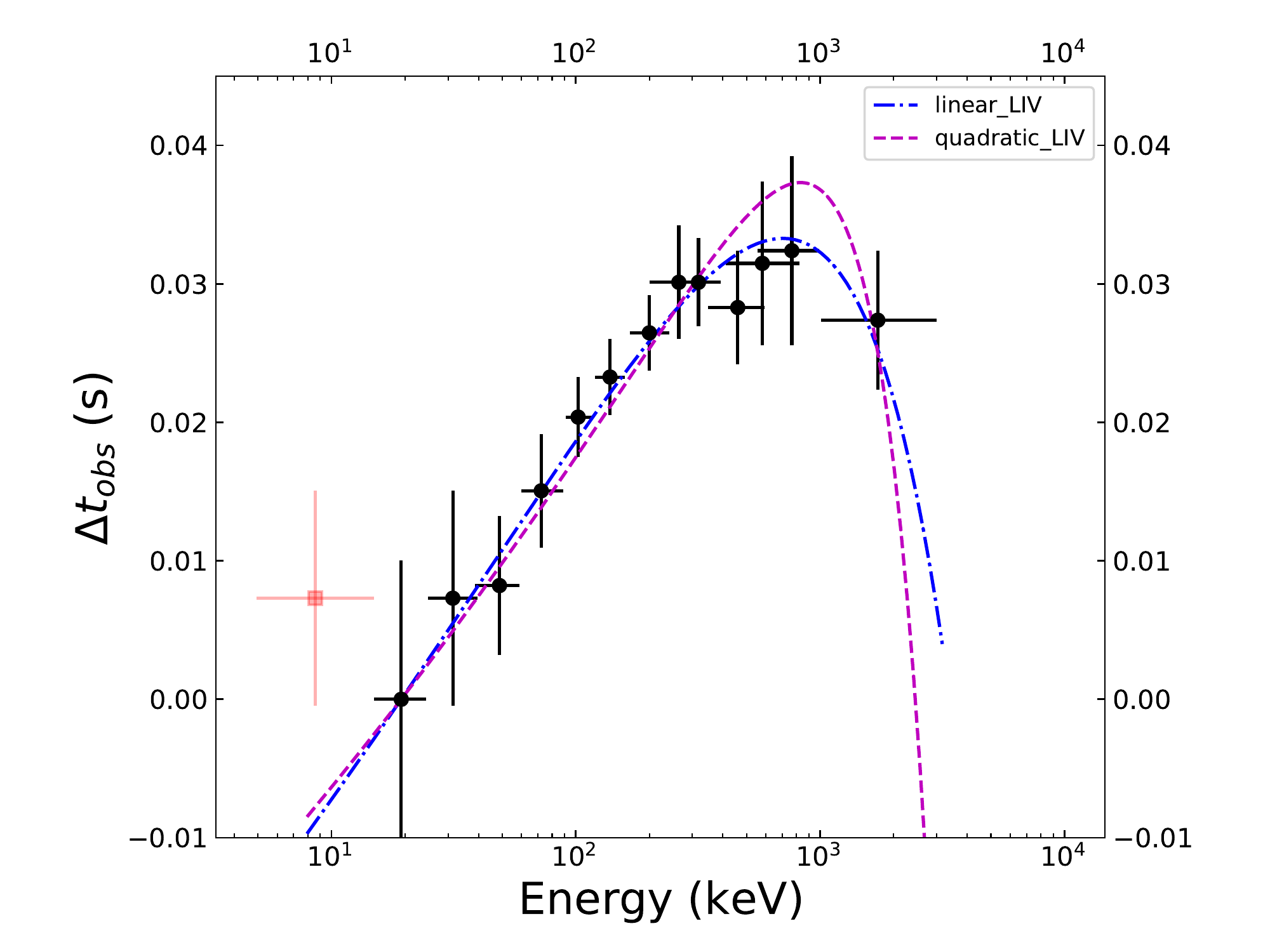}
\includegraphics[scale=0.28]{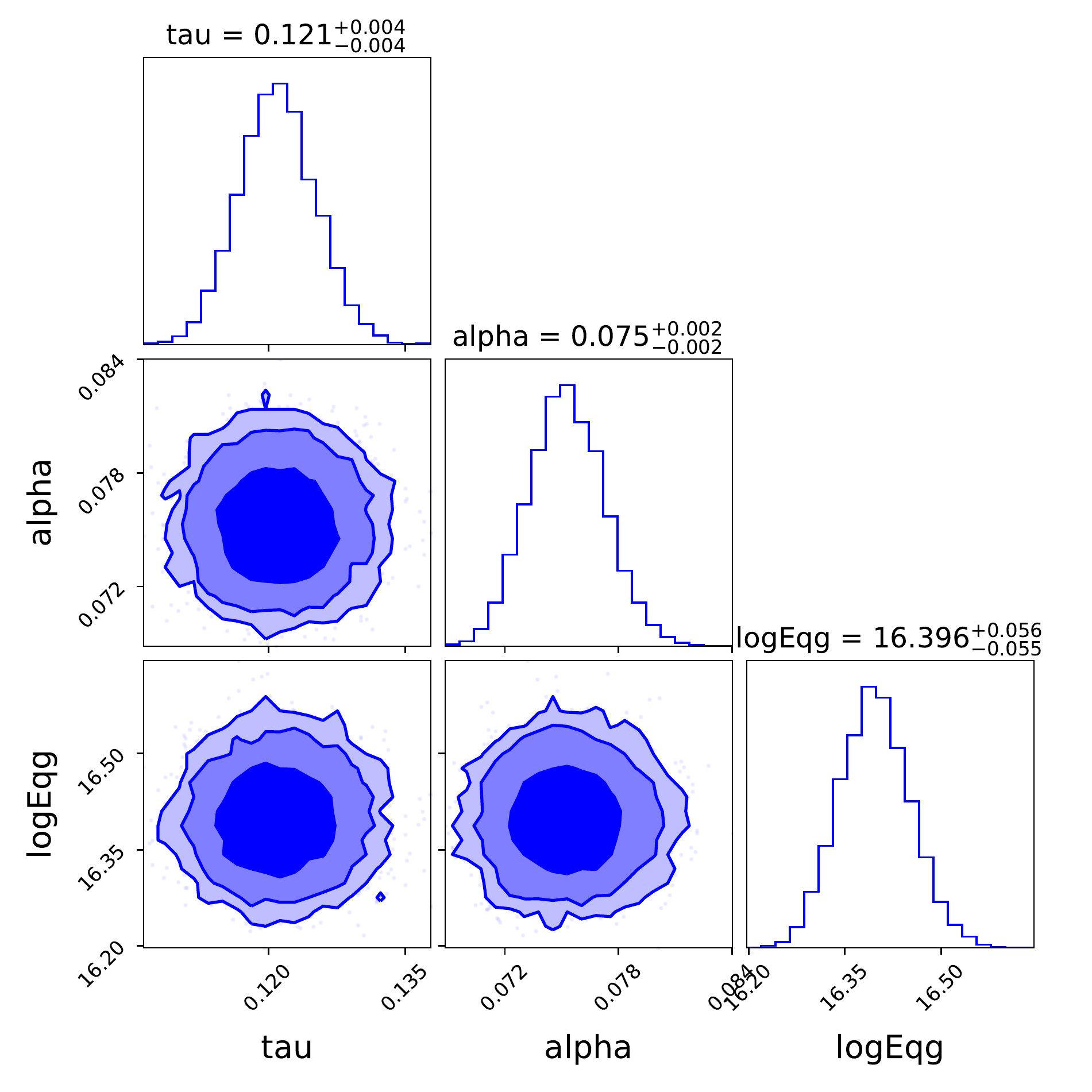}
\includegraphics[scale=0.28]{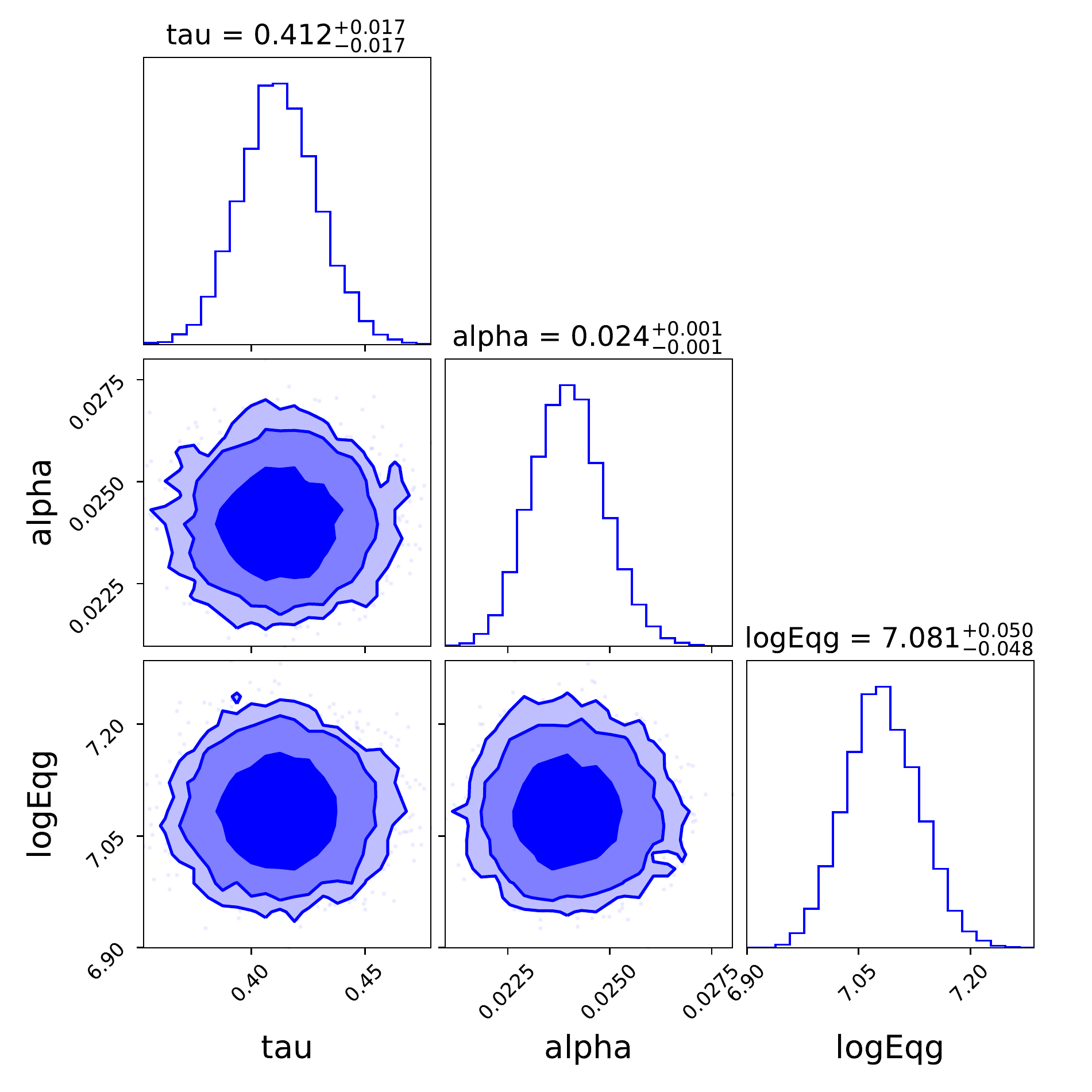}
\caption{The same as Fig.\ref{pusle2} but for the observed spectral lags in the total emission episode of GRB 190530A. }
\label{ptotal}
\end{figure}

Although most applications of Bayesian inference for parameter estimation and model selection in astrophysics involve the use of Monte Carlo techniques such as Markov Chain Monte Carlo (MCMC) and nested sampling (see \cite{Saha94,Kucukelbir16,Hogg18}). However, these techniques are time consuming and their convergence to the posterior could be difficult to determine. Whereas variational inference converts the inference problem into an optimization problem by approximating the posterior from a known family of distributions and using Kullback-Leibler divergence to characterize the difference, thus it can provide an approximate posterior for Bayesian inference faster than simple MCMC \citep{Kucukelbir16,Gunapati2022}. Therefore, in this paper, with the lag-energy data measured in the three episodes of GRB 190530A mentioned above, we perform variational inference to determine the free parameters ($\tau$, $\alpha$, and $ E_{\rm {QG,n}}$) in  Eqs. of (\ref{eq:int}) and (\ref{eq:liv}) based on the PyMC3 package \citep{Salvatier16}. Our analysis results are shown in Figs. of (\ref{pusle2}), (\ref{pusle3}), and (\ref{ptotal}), respectively. Their corresponding best parameters are also listed in Table (\ref{tab:results}). Noticing that, the red square points, as shown in Figs. of (\ref{pusle2}) and (\ref{ptotal}), are clearly outliers to our model, and now we do not have an accurate model for the outliers. So the data points are not included in our analysis. 

In all, we could come to the conclusions based on our fitting results: 

a) For the two LIV models, the three model parameters ($\tau$, $\alpha$, and $ E_{\rm {QG,n}}$) can be best constrained in the three emission episodes, respectively. Furthermore, there are no correlations between any pair-parameters for the three parameters based on their contours shown in Figs. of (\ref{pusle2}), (\ref{pusle3}), and (\ref{ptotal}).

b) Using the best-fit values of LIV model parameters  and their 1 $\sigma$ error seen in Table (\ref{tab:results}), we  obtain a 1 $\sigma$ lower limit on LIV as follows:

Second episode:
\begin{align*}
E_{QG,1} \geqslant 3.3 \times 10^{16} \text{ GeV} \hspace{0.35cm} &( n = 1 );\\ 
E_{QG,2} \geqslant 1.8 \times 10^{7} \text{ GeV} \hspace{0.35cm}	&( n = 2 ) 
\end{align*}

Third episode:
\begin{align*}
E_{QG,1} \geqslant 1.5 \times 10^{16} \text{ GeV} \hspace{0.35cm} &( n = 1 );\\
E_{QG,2} \geqslant 9.5 \times 10^{6} \text{ GeV} \hspace{0.35cm}	&( n = 2 ) 
\end{align*}

Total episode:
\begin{align*}
E_{QG,1} \geqslant 2.1 \times 10^{16} \text{ GeV} \hspace{0.35cm} &( n = 1 );\\
E_{QG,2} \geqslant 1.0 \times 10^{7} \text{ GeV} \hspace{0.35cm}	&( n = 2 ) 
\end{align*}

c) According to the qualitative criterion based on Jeffreys scale \citep{Jeffreys61}, with the resulting Bayes Factors (seen in Table (\ref{tab:results})), we argue that, when compared the two LIV models (linear ($n=1$) and quadratic ($n=2$)) with the observed spectral-lag data for the three episodes, respectively, any model could not significantly favoured by the data over another one. In other word, the two LIV models are comparable in the compatibility with the observed spectral-lag data.

However one would wonder that which the fitted model is correct? Traditionally, this is quantitatively addressed by Goodness of Fit (GOF) measures such as the reduced $\chi^2$ for normal distributions, for testing whether the fitted model is correct. But often, the reduced $\chi^2$ does not provide a reliable method for assessing and comparing model fits because the two independent problems (detailed information please referred to \cite{Andrae10}): i) The number of degrees of freedom  is unknown, spectially for nonlinear models; ii) random noise in the data is uncerain, which impairs the usefulness of reduced $\chi^2$. 

As a alternative methods to the reduced $\chi^2$, here we adopt a test statistic of normalised residual, a more sophisticated and reliable methods, which are also applicable to nonlinear models.
For the true model having the true parameter values and a-priori known measurement errors, the statistic follows by a standard Gaussian distribution. For any other model, this is not true \citep{Andrae10}. 

\begin{figure}
\centering
\includegraphics[scale=0.28]{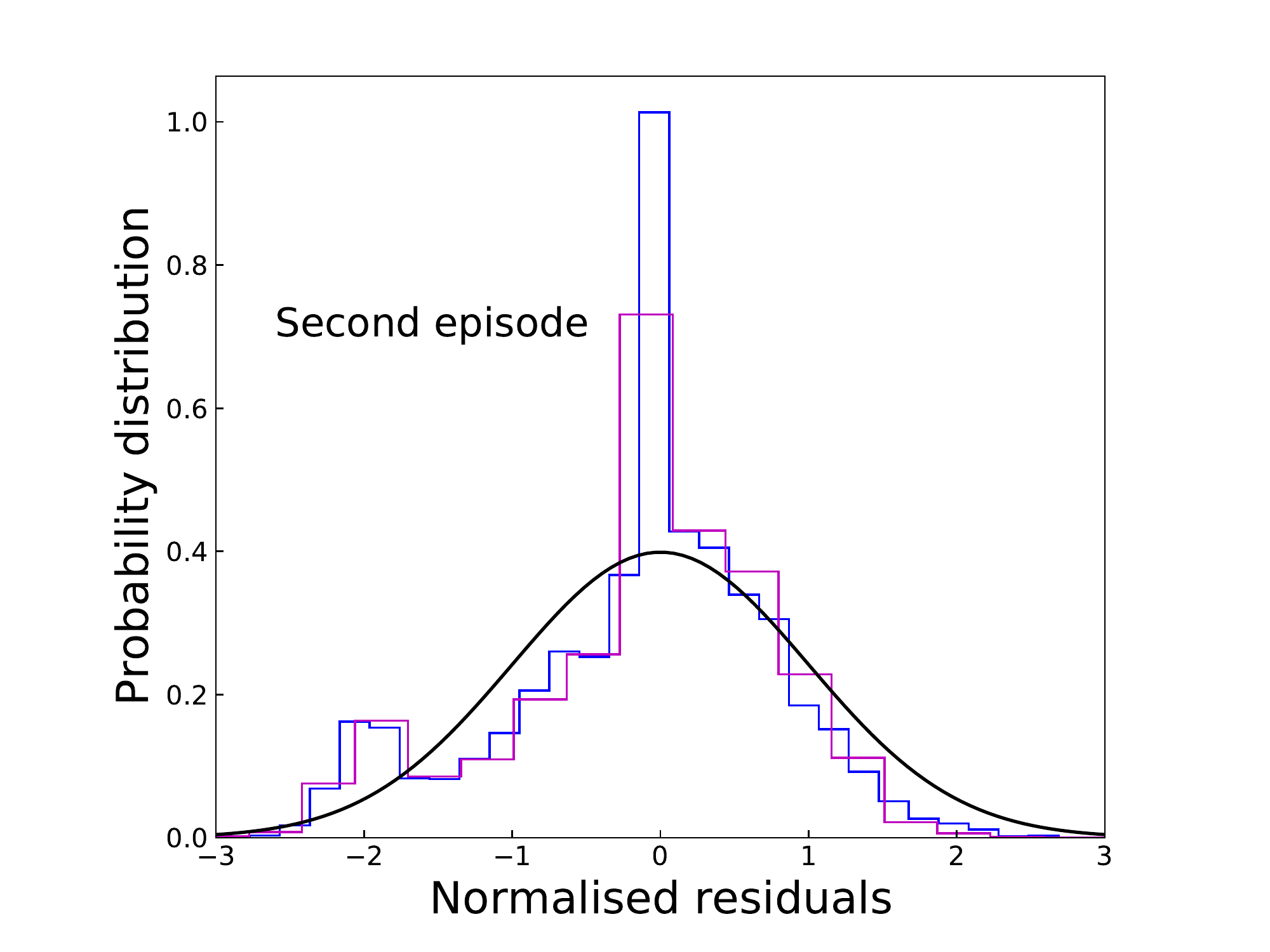}
\includegraphics[scale=0.28]{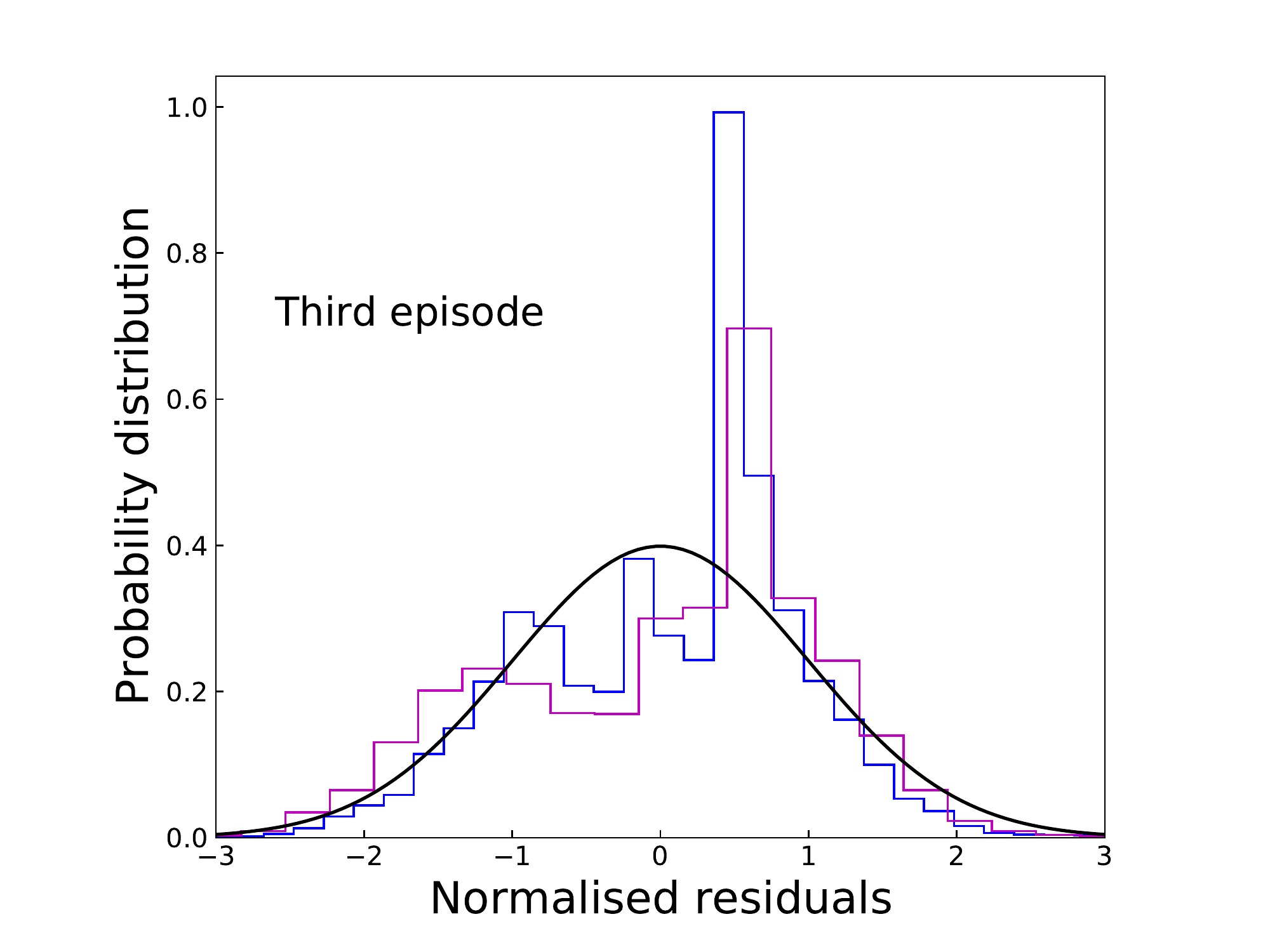}
\includegraphics[scale=0.28]{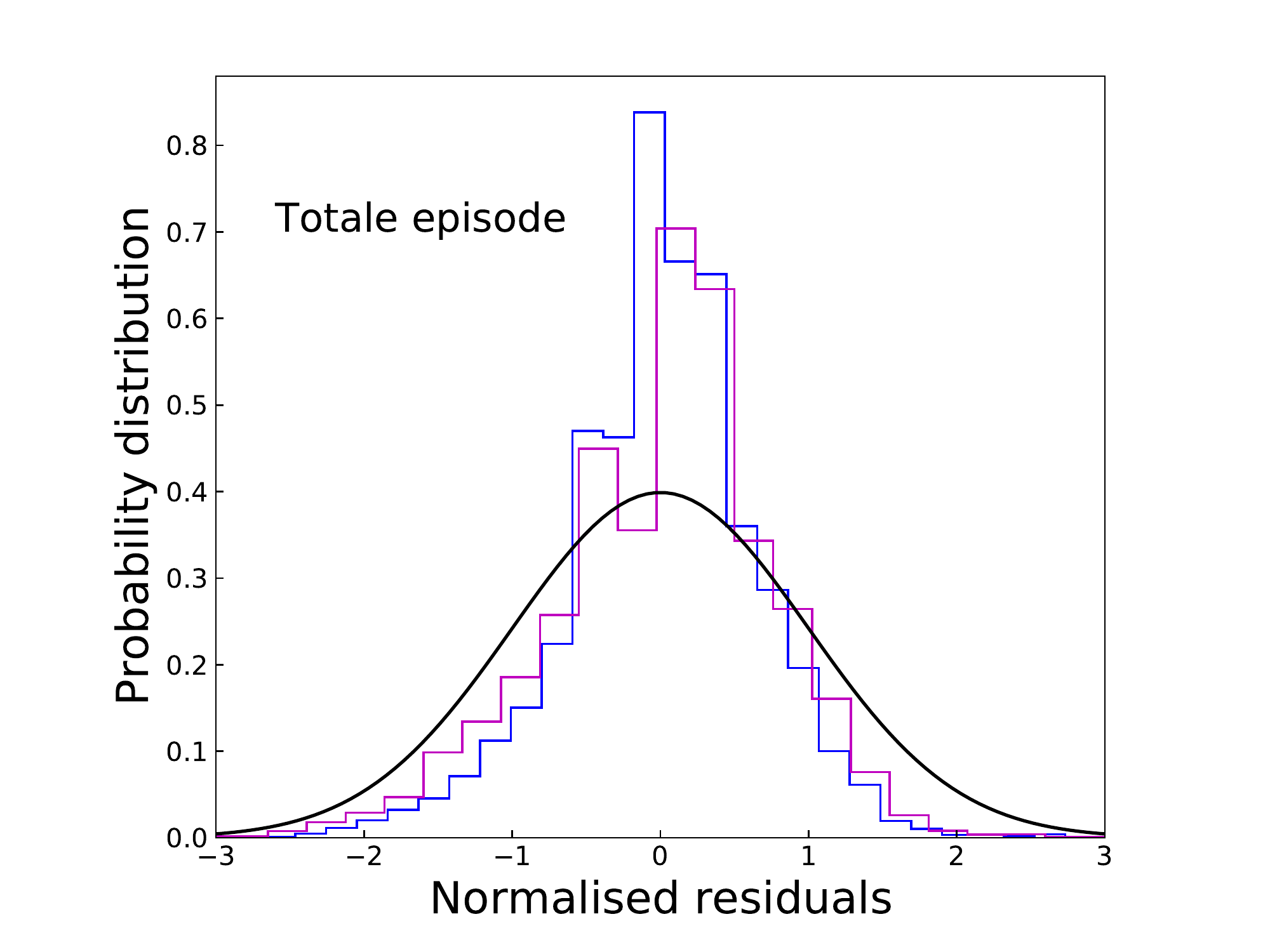}
\caption{Distributions of normalised residual between the observed spectral lags (from left to right pannels are for the data measured in the second, third episod and its total emission episode) and the LIV models for $n=1$ (blue histogram) and $n=2$ (red histogram), respectively.  For comparison we also show the standard Gaussian distribution (black histogram) as a referent one.}
\label{residual}
\end{figure}

By sampling from the Bayesian posterior distributions derived for the analysis above, we can calculate the normalised residuals between the LIV models and the observed spectral-lag data. Fig. (\ref{residual}) displays the distributions of their normalised residual derived from the three emission episodes, respectively. For every LIV model, we apply the KS-test to its normalised residual, and then the resulting p-values and its corresponding sigma levels\footnote{A p-value is the probability of the true model can generate residuals that agree with the standard Gaussian as badly as or worse than the actually observed ones. The smaller the p-value, the stronger the evidence that you should reject the null hypothesis. Sigma level reflect people's confidence that the result is not just down to chance. The higher the sigma level, the greater the confidence that we can reject the null hypothesis.} are also listed in the last two columns of Table (\ref{tab:results}). Thus one could find that, the significances of $n=1$ LIV and $n=2$ LIV model are 22.49$\sigma$ and 18.75$\sigma$ for the second episode, 24.72$\sigma$ and 31.09$\sigma$ for the third episode, and 33.31$\sigma$ and 23.15$\sigma$ for the total episode, respectively. Evidently, the null hypothesis (i.e., the fitted model is compatible with the observation) is rejected at a much high significance level for all the three emission episodes, which means that neither the linear ($n=1$) nor quadratic ($n=2$) LIV model is compatible with the observed spectral-lags measured in all the three emission episodes of GRB 190530A. Compared with the test statistic of the reduced $\chi^2$ applied to GRB~160625A\citep{Wei2017,Ganguly2017}, we here obtain a robust evidence that the LIV models are not compatible with the observations in \thisgrb.

\section{Case of spectral evolution induced observed spectral lags} \label{sec:spectral evolution}
Increasing evidence show that observed spectral lags measured in GRBs are directly related to both spectral evolutions and its spectral profiles. The energy-dependences of the observed spectral lag can be naturally reproduced by numerical simulations based on the phenomenological model with the observed spectral evolution pattern in an individual light curve pulse \citep{Lu2018, Du2019}. Being a smooth ‘‘ fast rise and exponential decay ’’ (FRED) pulse (see \cite{KL2003,Gunapati2022}), a clear spectral evolutions are observed in the second episode of \thisgrb (also see the left panel of Fig. \ref{LC_spectra}). Therefore, as an example, in this section, we will test where the turnover behavior seen in the pulse can be reproduced based on the phenomenological model, written as 
\begin{equation} \label{flux}
f(t, E)=I(t)\phi(E,t),
\end{equation}
where $f(t,E)$ is the flux density of a pulse observed at a given time $t$ and photon energy $E$, $I(t)$ and $\phi(E,t)$ are the intensity and normalized emission spectrum, respectively. The intensity $I(t)$ is described with an empirical pulse model \citep{KL2003}, which can be rewritten as follows
\begin{equation}\label{KRL}
I(t)=I_{\rm m}(\frac{t+t_0}{t_{\rm m}+t_0})^r[\frac{d}{d+r}+\frac{r}{d+r}(\frac{t+t_0}{t_{\rm m}+t_0})^{(r+1)}]^{-\frac{r+d}{r+1}},
\end{equation}
where $t_{0}$ measures the offset of the start of the pulse reference to the trigger time; $t_{\rm m}$ is the time of the pulse's maximum flux, $I_{\rm m}$; $r$ and $d$ are the power-law rise and decay indexes, respectively. 

Firstly, to capture the characteristic of a light curve pulse profile , we employ Eq.(\ref{KRL}) to fit the second episode of \thisgrb, and then obtain the best-fitted parameters as follows: ($f_{\rm m}$, $t_{\rm 0}$, $t_{\rm m}$, $r$, $d$)=($794.95\pm 8.609$, $6.000\pm 5.460$, $9.021\pm 0.049$, $9.283\pm 2.533$, $2.070\pm 0.353$), whose best-fit line is also shown as a red-dash one in the left panel of Fig. (\ref{LC_spectra}).

A three-segment broken power-law model (\sw{bkn2pow}\footnote{\url{https://heasarc.gsfc.nasa.gov/xanadu/xspec/manual/node141.html}}), has been adopted to perform time-resolved spectral analysis of \thisgrb by \cite{Gupta2022}. The \sw{bkn2pow} is 
\begin{equation}
N(\rm E)=  
\begin{cases}
K E^{-\Gamma_1}, \hspace{0.5cm} if E \leq E_{\rm {b,1}}   \\
K E^{\Gamma_2 -\Gamma_1}_{E_{\rm {b,1}}} (E/1keV)^{-\Gamma_2},  if E_{\rm {b,1}} \leq  E \leq E_{\rm {b,2}} \\
K E^{\Gamma_2 -\Gamma_1}_{E_{\rm {b,1}}} E^{\Gamma_3 -\Gamma_2}_{E_{\rm {b,2}}} (E/1keV)^{-\Gamma_3}, if E_{\rm {b,2}} \leq  E  
\end{cases}
\label{bkn2pow}
\end{equation}
where the two spectral breaks ($E_{\rm {b,1}}$ and $E_{\rm {b,2}}$) for the time-resolved spectra in the second episode of \thisgrb can be find in Table (A.6) of \citep{Gupta2022}, which are also shown in the left panel of Fig. (\ref{LC_spectra}). For a comparison, its' corresponding light curve (black solid) from the brightest detecter NaI 0 is also over-plot in the same figure. One could find from the figure that, for the $E_{\rm break, 2}$, it follows by an `intensity-tracking' evolution trend throught the whole pulse; whereas, the $E_{\rm break, 1}$ follows a `hard-to-soft'-dominated evolution (as argured in \cite{Lu2018}) based on the fact that, its' value increases from 18 keV to 84 kev in the rising phase, whereas decays from 167 kev decay to 15 keV in the decay phase (please referred to the black solid circle data points in the figure).

Also, to feature the spectral evolution of the two spectral breaks, we employ the following two functions to fit the observed data:
\begin{equation}
\log E_{\rm {b,2}}  \propto 
\begin{cases}
k_{\rm {2,r}} \log (t-t_{0}), \hspace{0.5cm} if \quad (t-t_{0}) \leq (t_{\rm m}-t_{0}) \\
k_{\rm {2,d}} \log (t-t_{\rm m}) \hspace{0.5cm}  if \quad  (t-t_{\rm m}) > 0
\end{cases}
\label{tracking}
\end{equation}
for ``tracking" evolution of the $E_{\rm {b,2}}$, and 
\begin{equation} \label{H2S}
\log E_{\rm {b,1}}\propto k_{\rm {1,d}} \log (t-t_{\rm m})  \quad if \quad  (t-t_{\rm m}) > 0
\end{equation}
for ``Hard-to-soft"-dominated evolution of $E_{\rm {b,1}}$. Thus we obtain that, $k_{\rm {2,r}}=1.01 \pm 0.09$, $k_{\rm {2,d}}=-1.34 \pm 0.08$, $k_{\rm {1,d}}=-1.64 \pm 0.40$. The best fitting results for the $E_{\rm {b,1}}$ (black dash line) and $E_{\rm {b,2}}$ (blue dash-dotted line) are also shown in the left panel of Fig. (\ref{LC_spectra}).

Based on the phenomenological model of Equation (\ref{flux}) with the observed features for the two spectral breaks evolutions, described by Equations of (\ref{tracking}) and (\ref{H2S}), together with constant power-law indices\footnote{Spectral analysis \citep{Gupta2022} shows that the values of $\Gamma_{\rm {1,2}}$ are consistent with the power-law indices expected for synchrotron emission. Therefore we will adopt fixed values for the three power-law indices in our numerical simulation, i.e., $\Gamma_{\rm {1}}=2/3$, $\Gamma_{\rm {2}}=3/2$, and $\Gamma_{\rm {3}}=2.5$.}, we numerically calculate the spectral lags, which are estimated based on the discrepancy of
light curves' peak time ($\Delta t_{\rm p}$), and then the results are plotted in the right panel of Fig.(\ref{LC_spectra}). One could find that our numerical spectral lags can follow well the observation (black squares) in term of its evolutionary trend. Especially, a turnover also appears in the energy of $E_{\rm {obs}}\sim 1\times10^3$ keV. Admittedly, there exists a difference in absolute values of the spectral lags, as seen in the right panel of Fig.(\ref{LC_spectra}). This may be the cause that, firstly, the spectral lags are sensitive to the spectral evolution and its spectral shape \citep{Lu2018, Du2019}, and a simple model is used in our simultions; Secondly, different analysis methods (here we have measured a lag by $\Delta t_{\rm p}$) may result in a systematic error in the measurement of the spectral lags. At the same time, one could find that there also exists a larger discrepancy between the observation and our simulation in the low-energy end (the first two data points in the right panel of Fig.(\ref{LC_spectra})). Although, in our simulations, we have tried to adopt different evolution patterns of the $E_{\rm {b,1}}$ in the rising phase of the pulse, such as ``soft-to-Hard" evolutions with different values of the $k_{1}$ in Equation of (\ref{H2S}), or no spectral evolution, the issue of the discrepancy could not be resolved. However, we have also noticed that there are larger uncertainties in the spectral lags for the first two time bins being with weak signals. Whether this observation is reality or not needs to be confirmed by more observations in the future. If it is the case, it means that a reverse evolutionary trend in the spectral lags takes place at the low-energy end. So its physical origins need to be investigated further. 

In all, our simulations by a simple hypothesis with Equation (\ref{flux}) can duplicate the observation results above, which may indicate that the difference of the light curve peaks in different energy band, which is known as the spectral lag, results from the time evolution of the spectral break ($E_{\rm {b,1}}$ and $E_{\rm {b,1}}$) across energy bands (e.g., \cite{Ukwatta12}), which will result in the difference of the light curve peaks when one integrates over different energy bands.
 
\section{DISCUSSION and SUMMARY} \label{sec:diss}
Our model evaluation above shows that neither the linear nor quadradic model is compatible with the current observations, although the quantum gravity scale, either $E_{\rm {QG,1}}$ or $E_{\rm {QG,2}}$ can be constrained well within a small error.

Furthermore, assuming that a positive-to-negative spectral lag transition can be correctly described by Equation (\ref{eq:sum}), we can derive the LIV-induced observed time delay between two photons emitted simultaneously from GRB 190530A based on Equation (\ref{eq:liv}) with the best parameters in Table \ref{tab:results}. For example, if the two photons with $E_l$=$1\times10^3$ \meV and $E_h$=$5\times10^3$ \meV are adopted, we get,
\begin{align*}
\Delta t_{\rm {LIV},1} = 0.041 \pm 0.015 \hspace{0.05cm} \text{s} \hspace{0.35cm} &( n = 1 ),\\
\Delta t_{\rm {LIV},2} = 0.047 \pm 0.027 \hspace{0.05cm} \text{s} \hspace{0.35cm}&( n = 2 ),
\end{align*}
for second episode,  
\begin{align*}
\Delta t_{\rm {LIV},1} = 0.115 \pm 0.009 \hspace{0.05cm}\text{s} \hspace{0.35cm} &( n = 1 )\\
\Delta t_{\rm {LIV},2} = 0.238 \pm 0.038 \hspace{0.05cm}\text{s}	\hspace{0.35cm}&( n = 2 ),  
\end{align*}
for third episode, and 
\begin{align*}
\Delta t_{\rm {LIV},1} = 0.078 \pm 0.010 \hspace{0.05cm}\text{s} \hspace{0.35cm} &( n = 1 )\\
\Delta t_{\rm {LIV},2} = 0.193 \pm 0.039 \hspace{0.05cm}\text{s}	\hspace{0.35cm}&( n = 2 ), \hspace{0.35cm} 
\end{align*}
for total emission episode. The results show that, the two photons have different time delays induced by LIV effect, This enhances the evidence that these LIV models do not fit the observed data. 

To search for possible LIV signatures, it is crucial to correctly evaluate the energy-dependent source intrinsic time-lag. It is generally believed that an individual shock episode gives rise to a pulse, and random superposition of many such pulses results in the observed complexity of GRB light curves. As a result, \cite{Hakkila08} suggested that the spectral lag is better defined using individual pulses rather than the whole burst light curve profile. Thus both the temporal and spectral behaviors of a pulse may be related the dynamics of the GRB, the electron acceleration, and the radiation mechanism (e.g.,\cite{Kobayashi97,Dermer98,Daigne98,Ramirez00}), which also are correlated to the observed spectral lags. Therefore the intrinsic spectral lags measured in different emission episodes even in the same GRB should be different (e.g.,\cite{Page07,Lu2018}). As we have found that there are different intrinsic time-lags in the three different emission episodes in \thisgrb. Again, it is worth mentioning that there is not any physics behind Eq. (\ref{eq:int}) firstly invoked by \cite{Wei2017}, and its correctness needs to be tested by more observations. So one should search for such a possible LIV signature in a smooth, well-separated pulse, in which the intrinsic spectral lag must be well known in advance. This is worthy of further investigation.

Based on the relativistic jet shell model constructed in \cite{Uhm16} with a cutoff power-law (CPL) radiation spectrum or a Band function spectrum \citep{Band93}, \cite{Du2019} found that the spectral lag monotonically increases with photon energy and finally saturates at a higher energy, which is consistent with the most observations (e.g., \citealp{Lu06,Lu2018}). Interestingly, when replacing with a Bandcut spectrum \citep{Tang15}, they got a turn-over from the positive lags to negative lags in the high-energy range. Next, based on a phenomenological model, which characterises the profile of the second pulse in GRB~160625B and it' corresponding Bandcut spectral evolutions \citep{Du2019}, they furtherly comfirmed that a turnover does take place at the high energy $E_{\rm {obs}}$ $\geq$ 40 MeV, the same as the observations. 

\begin{figure}
\centering
\includegraphics[scale=0.4]{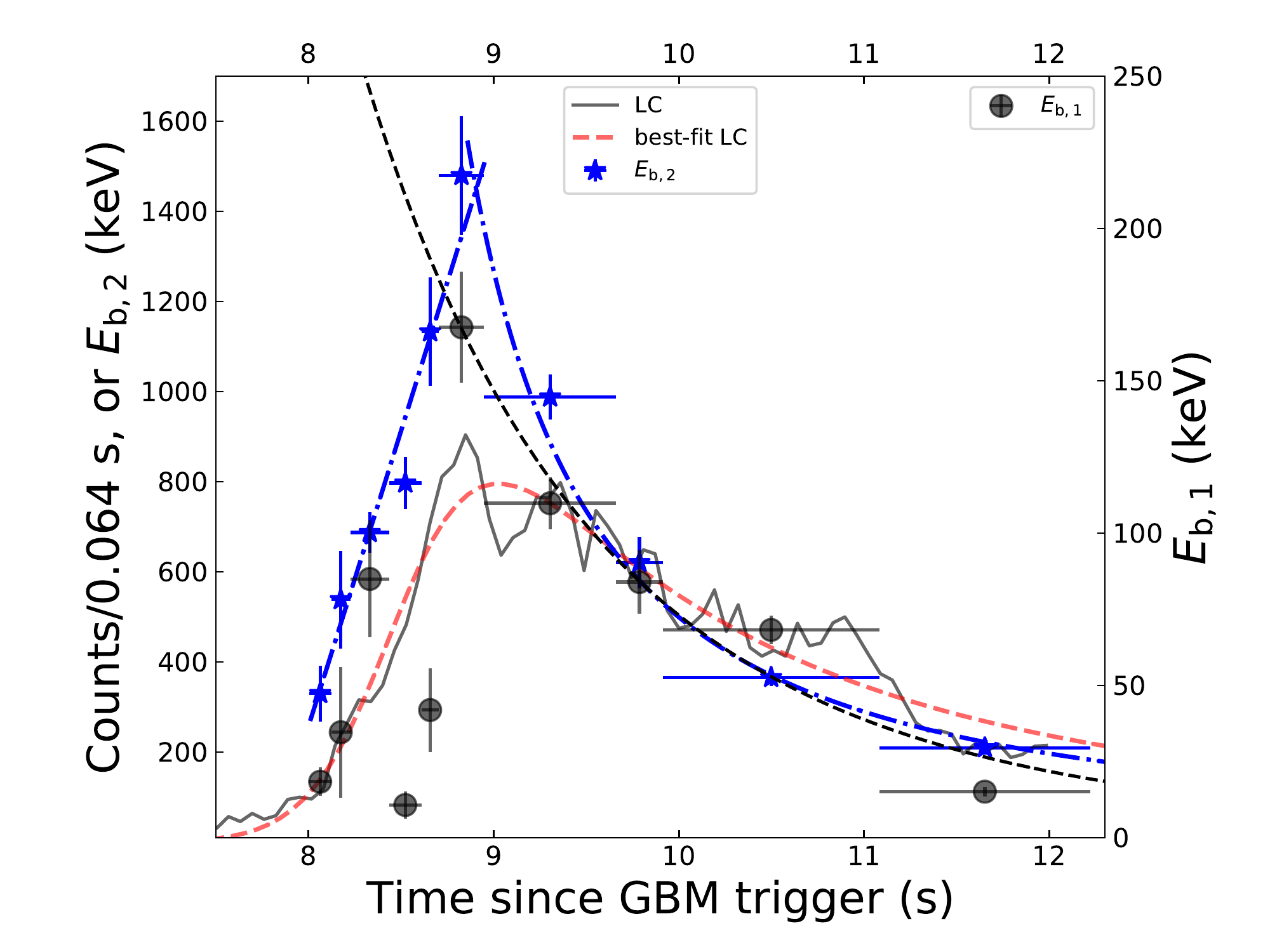}
\includegraphics[scale=0.4]{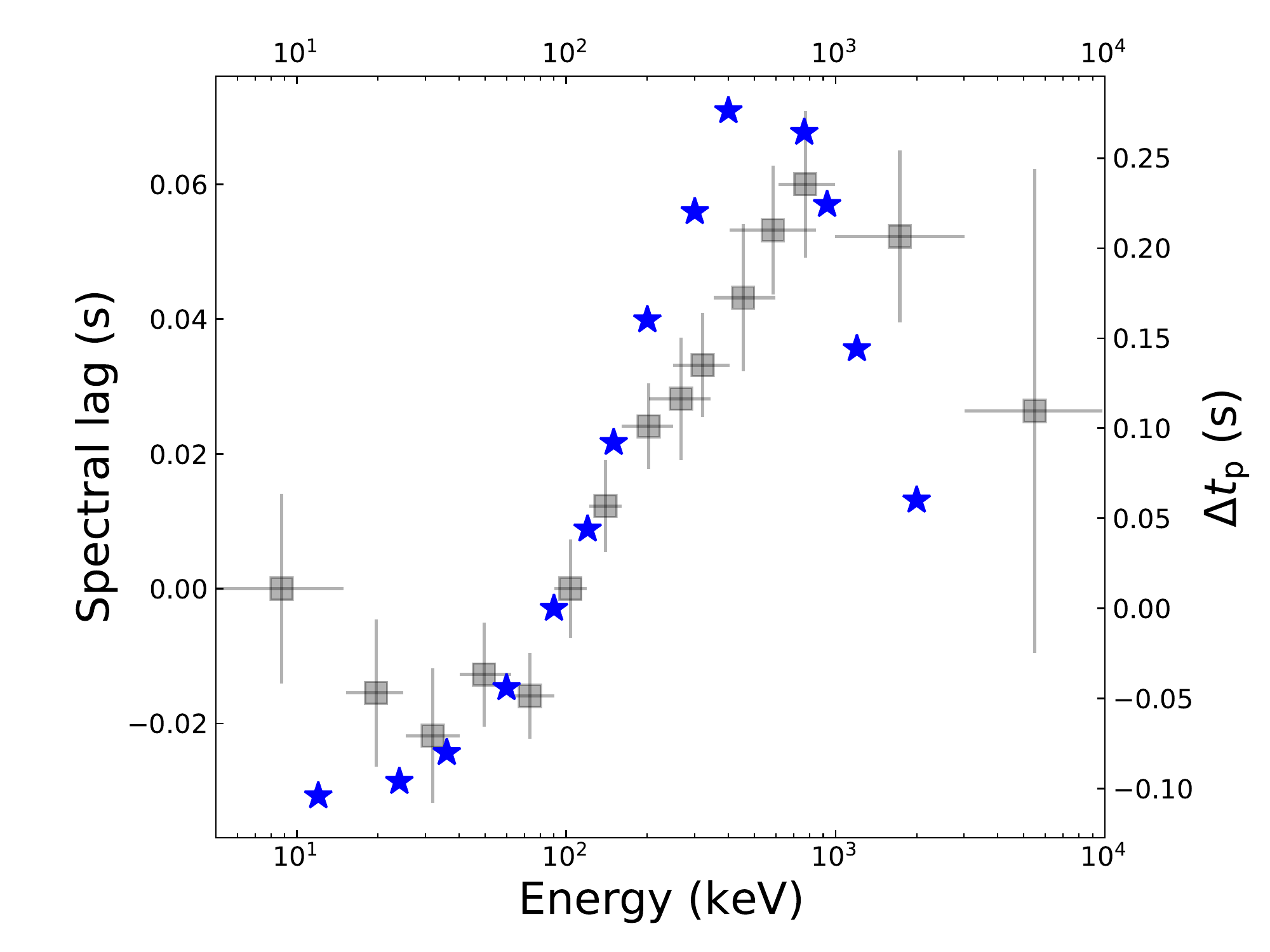}
\caption{Left panel: Demonstrations of the evolution pattern of the two spectral breaks, $E_{\rm break,1}$ (black solid circles, right y-axis) and $E_{\rm break,2}$ (blue stars, left y-axis), measured in the second pulse of \thisgrb (black solid line, left y-axis). The black dash and the blue dash-doted lines represent the best fitting results of $E_{\rm break,1}$ and $E_{\rm break,2}$ with Equations of (\ref{H2S}) and (\ref{tracking}), respectively. Data for $E_{\rm break,1}$ and $E_{\rm break,2}$ comes from Table (A6) in \cite{Gunapati2022}. Right panel: energy-dependence of the observed spectral lags (blue star, right y-axis) based on our simulations with the feature of the spectral break evolutions as seen in the left panel, and the spectral indexes of $\Gamma_1 =-2/3$, $\Gamma_2 =-3/2$ and  $\Gamma_3 =-2.4$ are adopted in the simulations. The spectral lag relative to the (90,120) keV channel. For comparison we also show the observed spectral lags (black square, left y-axis, data from Figure (A2) in \cite{Gunapati2022}). }
\label{LC_spectra}
\end{figure}

Here we copy the same method adopted in \cite{Du2019}, but for a \sw{bkn2pow} spectrum seen in \thisgrb, and find that our numerical results can reproduce the observed spectral lag (see the right panel of Fig. (\ref{LC_spectra})). The fact shows that such a positive-to-negative spectral lag transition can be reproduced well in the case of spectral evolutions, in which don't need introduce additional LIV-induced observed spectral lag.

Whether the turn-over results from LIV lags or from spectral evolutions? it needs more observations to furtherly comfirm. Anyway, to search for LIV signatures, one must carefully take care of the impact of the intrinsic time-lag on the observed spectral time-lags.  

\clearpage

\begin{acknowledgments}

We acknowledge the use of the Fermi archive’s public data. This work is supported by the Guangxi Science Foundation (grant No. 2018GXNSFDA281033) and the National Natural Science Foundation of China (grant No. U1938106). 
\end{acknowledgments}

\vspace{5mm}

\software{PyMC3 \citep{Salvatier16}}

\bibliography{refs}
\begin{table}
\caption{The best-fit values of the the two LIV model parameters, as well as their 1 $\sigma$ errors, and the p-values and sigma levels for statistical significance test.}
\label{tab:results}
\begin{center}
\begin{tabular}{|c |c | c | c | c | c | c | c |}
\hline
\textbf{Emission interval} & \textbf{$n^{a}$} & \textbf{$\tau$ (s) } & \textbf{$\alpha$ }& \textbf{$\log10(E_{\rm {QG,n}}/Gev)$} &\textbf{$BF_{12}^{b}$} & \textbf{p-value} & \textbf{Sigma level}  \\
\hline
Second episode & n=1 & 0.235$^{+0.011}_{-0.011}$ &0.074$^{+0.003}_{-0.003}$ & 16.668$^{+0.122}_{-0.145}$ &  & 5.362E-112 & $22.49\sigma$\\
  & n=2 & 0.295$^{+0.015}_{-0.015}$ &0.059$^{+0.002}_{-0.002}$ & 7.364$^{+0.088}_{-0.099}$ & 0.128 & 1.828E-78 & $18.75\sigma$\\
\hline
Third episode & n=1 & 0.061$^{+0.002}_{-0.002}$ &0.126$^{+0.002}_{-0.002}$ & 16.220$^{+0.036}_{-0.033}$ &  & 2.725E-134 & $24.72\sigma$\\
  & n=2 & 0.364$^{+0.012}_{-0.011}$ &0.026$^{+0.001}_{-0.001}$ & 7.014$^{+0.035}_{-0.033}$ & 3.734 & 1.015E-211 & $31.09\sigma$\\
\hline
Total interval & n=1 & 0.112$^{+0.004}_{-0.004}$ &0.080$^{+0.002}_{-0.002}$ & 16.387$^{+0.056}_{-0.054}$ &  & 1.111E-242 & $33.31\sigma$\\
  & n=2 & 0.154$^{+0.006}_{-0.006}$ &0.056$^{+0.002}_{-0.002}$ & 7.059$^{+0.046}_{-0.043}$ & 4.095 & 1.332E-118 & $23.15\sigma$\\
\hline
\end{tabular}
\end{center}
\indent \indent $\color{blue}^a${\small \textit{LIV model: linear (n=1) order and quadratic (n=2) order}}
\indent \indent \indent \indent $\color{blue}^b${\small \textit{Bayes Factor.}}\\

\end{table}



\end{document}